\documentstyle[multicol,aps,epsf,here]{revtex} 
\tighten
\newcommand{\PostScript}[7]{
\begin{figure}[H]
\begin{center} 
\leavevmode
\epsfysize=#1cm
\vspace{#2cm}
\epsfbox{#3}
\par
\parbox{#5cm}{
\vspace{#4cm}
\caption[figure]{\renewcommand{\baselinestretch}{1} \small \normalsize #6}
\label{#7}}
\end{center}
\end{figure}
}

\begin{document}

\title{On the formation of Wigner molecules in small quantum dots}

\author{S.M. Reimann, M. Koskinen, and M. Manninen}      
\address{\it Department of Physics, University of Jyv\"askyl\"a,
FIN-40351 Jyv\"askyl\"a, Finland}

\medskip

\maketitle

\begin{abstract}
It was recently argued that in small quantum dots the electrons 
could crystallize at much higher densities than in the 
infinite two-dimensional electron gas. 
We compare predictions that the onset of spin polarization and
the formation of Wigner molecules occurs at a density parameter 
$r_s\approx 4~a_B^*$ to the results of a straight-forward diagonalization 
of the Hamiltonian matrix. 
\end{abstract}
\begin{multicols}{2}
\narrowtext

\section{Introduction}

If the number of electrons artificially confined on a 
quasi two-dimensional electron island (made for example 
in a semiconductor heterostructure) is very large, many  
properties of such a so-called ``quantum dot'' 
or ``artificial atom'' ~\cite{ashoori} can be described from 
what is known about the limit of the infinite (two-dimensional) 
electron gas (2DEG). Until now, 
most experiments~\cite{kouwenhoven:review} were performed at 
electron densities which are slightly below the equilibrium density 
of the 2DEG. The liquid-like properties then still dominate. 
For only a few trapped particles (as experimentally realized 
in vertical quantum dots~\cite{tarucha}), 
pronounced addition energy maxima as a consequence of shell structure 
and aligned spins in the mid-shell regions due to Hund's rules 
were observed in close analogy to atomic physics~\cite{tarucha}. 
Even the simplest picture of $N$ non-interacting particles 
in a two-dimensional harmonic trap could explain many features 
of the conductance spectra.
For larger systems, mean field approaches like 
Hartree-Fock~\cite{koonin,landman} or density functional 
methods~\cite{stopa,koskinen,steffens,lee,reimann,austing,hirose} 
were applied.  
In the small-$N$ limit, much theoretical work focused on exact diagonalization 
techniques~\cite{pfann,exact}. This approach was mostly used 
for dots in magnetic fields, where correlations become increasingly important 
with field strengths. It was particularly successful in the
(integer and fractional) quantum Hall regime where one can restrict 
the basis set to the lowest spin-polarized Landau level~\cite{laughlin,girvin}.
Quantum Monte Carlo (QMC)~\cite{bolton2,egger,pederiva,harju}
methods provide alternative approaches yielding energies whose accuracy 
reaches that of exact diagonalizations.

When the electron density is lowered and the Coulomb energy 
increases relative to the kinetic energy, correlations 
begin to strongly dominate the electronic structure also in the absence 
of magnetic fields.
For densities smaller than a certain critical value a Wigner 
crystal~\cite{wigner} will be formed, 
in which the Coulomb interaction distributes the 
single electrons classically on a lattice.
For the homogeneous two-dimensional electron gas, such crystallization is 
expected at very low densities. Monte-Carlo calculations indicate 
that in the 2D bulk a transition to a Wigner crystal-like state, 
preceeded by a transition to a polarized phase~\cite{rapisarda}, occurs 
only at densities corresponding to Wigner-Seitz radii 
$r_{s,2D}~>~37~a_B^*$~\cite{tanatar}
(the density $n_0$ and $r_{s,2D}$ are related by $n_0=1/(\pi r_{s,2D}^2)$),
whereas in 3D the classical limit lies as high as 
$r_{s,3D}=100~a_B^*$~\cite{ceperley}.
(In the following, for simplicity we write $r_{s,2D}=r_s$.)
Chui and Tanatar~\cite{chui} found that in 2D systems 
without translational invariance the critical 
density parameter for a fluid-solid transition is shifted 
to a considerably smaller value ($r_s~\approx~7.5~a_B^*$).
Could this be important for {\it finite} systems such as the above mentioned 
lateral or vertical semiconductor quantum dot structures?
This question was recently posed~\cite{landman,egger,creffield} and
it was argued that in finite systems confining only a few particles
localization would indeed occur at significantly higher densities than 
in the 2D bulk. In the Wigner limit the few electrons in the trap 
would distribute such that their electrostatic repulsion is minimized. 
The internal structure of the wave function of the many body system should 
then have the symmetry of the corresponding classical charge distribution. 
Wigner crystallization was found to be particularly pronounced in 
quantum dots with steep walls and polygonal geometry~\cite{creffield}.
Egger {\it et al.}~\cite{egger} have performed quantum Monte Carlo studies
using a multilevel blocking algorithm~\cite{mlb}.
For parabolic quantum dots with azimuthal symmetry
they reported that at a critical density of $r_s~=4~a_B^*$
the formation of Wigner molecule-like ground states should
become energetically favorable. Hartree-Fock (HF) methods by definition fail
to accurately describe the correlated regime. 
When performed in an unrestricted scheme, however,
spontaneous symmetry breaking and localization
in the spatial distribution of the electronic 
densities of quantum dots and lateral quantum 
dot molecules at $r_s\approx 3.5~a_B^*$ was attributed to the onset of 
Wigner crystallization~\cite{landman}.

In the present article we report numerically exact configuration 
interaction calculations. This method has a long history in quantum 
chemistry, and was applied to quantum dots by many different 
groups~\cite{exact}.
Much of the previous work, however, concentrated on the electronic structure 
in large magnetic fields where the electron gas is polarized.
Our purpose here is a comparison of the 
exact diagonalization results to the above mentioned recent predictions of 
localized states in the low-density limit and zero magnetic field. 
We first give a brief outline of the configuration 
interaction method and then turn to 
a discussion of the many-body spectra of a six-electron quantum dot 
at zero angular momentum as a function of the average electron density 
in the dot. Calculations for different $r_s$-values indicate that 
the ground state remains unpolarized.
At values of $r_s$ which are accessible to exact diagonalization techniques,
for a dot confining six electrons clear signals of formation of a 
Wigner molecule could not be observed. 
We conclude with a brief comparison to results of 
density functional theory (DFT) 
in the local density approximation (LSDA) describing 
the electronic ground state structures. 

\section{Method and Convergence}

Consider $N$ interacting electrons 
trapped in a circularly symmetric 
harmonic well $V(r) = m^* \omega _0 ^2 r^2 /2~$, where 
$r^2=x^2+y^2$. (In the quasi two-dimensional limit
one assumes that the confinement in $z$-direction is much stronger than 
in the $x$-$y$--plane. Then, only the lowest subband in $z$-direction 
is populated.)
We write for the Hamiltonian 
\begin{equation}
H=\sum _{i=1}^N \biggl [ {-\hbar ^2\over 2m^*} \nabla _i^2 + V(r_i)
\biggr ] + \sum _{i<j}^N {e^2\over 4\pi \varepsilon _0 \varepsilon} 
{1\over \mid {\bf r}_i - {\bf r}_j\mid }~.\label{ham}
\end{equation}
Here, $m^*$ and $\varepsilon $ are the effective mass and the 
dielectric constant. 
The calculations are done for different values of the density parameter $r_s$
which determines the average particle density in the dot, 
$n_0=1/(\pi r_s^2)$. The latter is approximated by setting the 
oscillator parameter $\omega _0^2 = e^2 / 
(4\pi \varepsilon _0 \varepsilon m r_s^3 \sqrt{N})$~\cite{koskinen}.
Throughout this paper we use effective atomic units
in which the length unit $a_B^*$
is a factor $\varepsilon /m^*$ times the Bohr
radius $a_B$, and the energy is given in effective Hartree, 
Ha$^*=~$Ha$\cdot(m^*/\varepsilon ^2)$. 
(For GaAs for example, $m^*=0.067m_e$ 
and $\varepsilon = 12.4$, for which the length and energy units then 
scale to $a_B^*=97.9{\hbox{~\AA}}$ and Ha$^*=11.9$~meV.)
To diagonalize the Hamiltonian, Eq.~(\ref{ham}),  
the spatial single-particle states of the Fock space are chosen
to be eigenstates of the two-dimensional harmonic oscillator 
with optimized oscillator parameter $\omega $.
In general, the electron-electron interaction tends to expand the system, 
and thus $\omega \ne \omega _0$. This effect becomes stronger 
with increasing $r_s$ and we used the (empirical) relation 
$\omega = \omega _0/\sqrt{r_s}$~.
The two-body matrix elements of the electron-electron interaction
are calculated using the addition
theorem for $1/\mid {\bf r}_1 - {\bf r}_2 \mid $~\cite{weissbluth}.
To set up the Fock states for diagonalization, we 
use eight lowest oscillator shells containing 36 states and
sample over the full space with a fixed number of spin down and spin 
up electrons, $N_{\downarrow} + N^{\uparrow } = N$.
From this sampling, only those Fock states 
with a given total orbital angular momentum 
and a configuration energy (corresponding to the sum of occupied 
single-particle energies) less than or equal to 
a specified cutoff energy $E_{\hbox{c}}$ 
are included (see Figure~\ref{f1} below for an example). 
The purpose was to select only the most important 
Fock states from the full basis, hereby reducing the matrix dimension
to a size $d\stackrel {<}{\sim} 10^5$.
To obtain all the eigenstates
we have to set $N_{\downarrow} = N_{\uparrow} =N/2$ for even particle 
numbers ($S_z=0$, all states with different total spin have this 
component), and analogously we would have 
$N_{\downarrow } = N_{\uparrow }\pm 1$ for odd numbers.
Once the active Fock states are at disposal, the Hamiltonian matrix is 
calculated. For Lanczos diagonalization we use the Arpack 
library~\cite{arpack}. Finally, the total spin of
each eigenvector is determined by calculating the expectation value
of the $\hat S^2$ operator. 

As mentioned above, setting up the Hamiltonian matrix from 
chosen Fock states and subsequent diagonalization only 
in principle yields an exact solution of the many-body problem.
For reasons of numerical feasibility it is necessary to truncate the set of
basis functions to be used in the diagonalization. 
One then has to make sure that convergence of the spectra is reached with 
respect to the cut-off. 
As the required matrix size increases rapidly with $N$,
computational expenses severely restrict the calculations to  
only the smallest systems at not too large values of $r_s$.
Thus, with increasing electron number or $r_s$,
the results become less accurate due to the restricted number of 
basis states that can be included in the calculations. 
(The fact that the ground and excited states 
are the closer in energy the larger the particle number $N$ 
imposes an additional difficulty.)
\PostScript{6}{0}{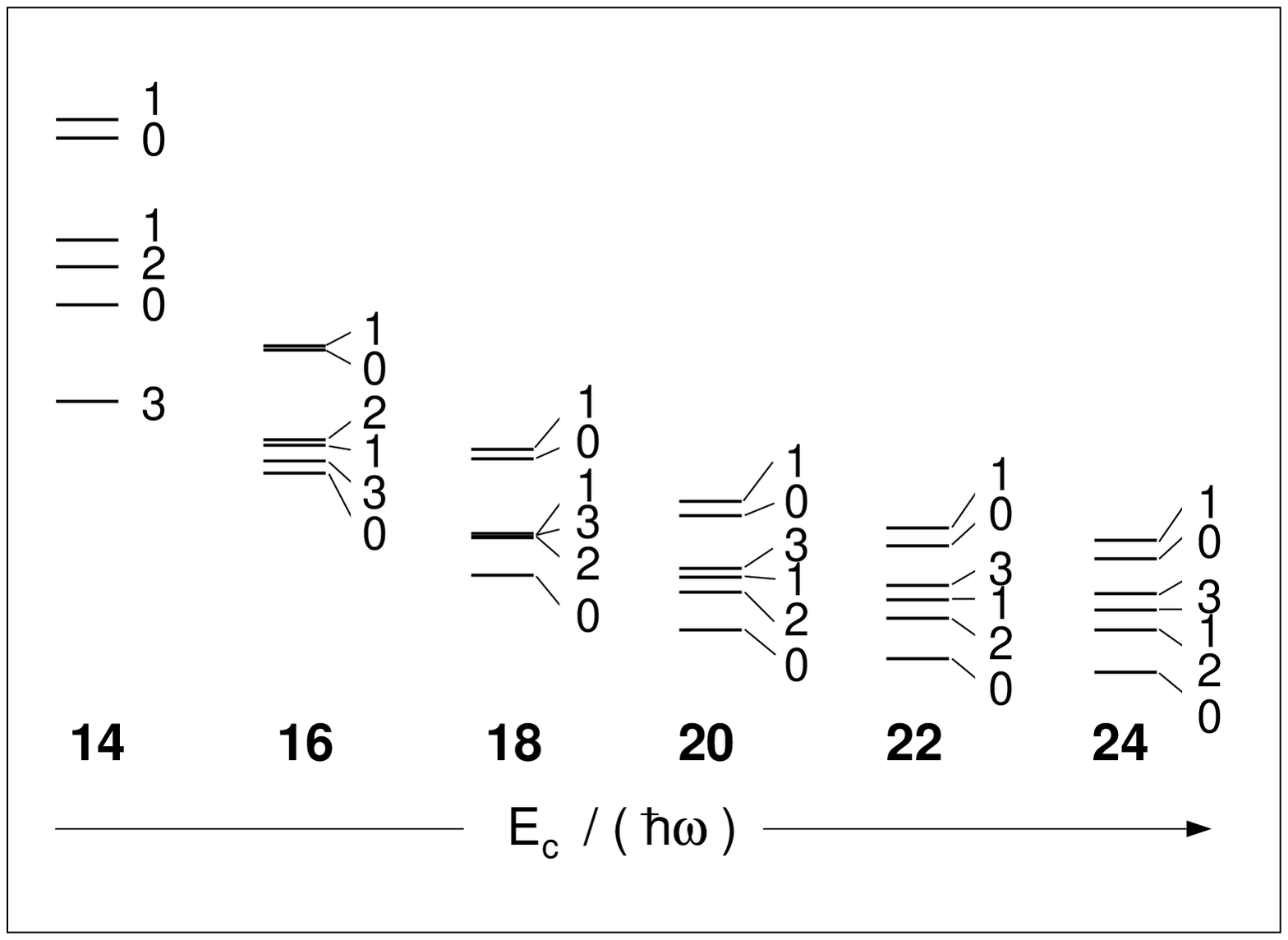}{0.5}{14}{\normalsize 
Convergence of the many-body spectra of $N=6$ 
electrons in a harmonic trap at a density corresponding to $r_s=4~a_B^*$ 
as a function of the cut-off energy $E_{c}/(\hbar \omega)$.
Shown are the 6 lowest states with $L=0$, 
together with the spin $S$ of each state.}{f1}
For a quantum dot confining $N=6$ electrons at a density 
corresponding to $r_s=4~a_B^*$ (the largest value of $r_s$ we found 
accessible within the calculational scheme used here),
Figure~\ref{f1} shows the convergence of the many-body spectra
as a function of the cut-off energy $E_{c}$.
The lowest possible Fock state for six electrons has two particles 
in the state $|n,l \rangle = |0,0 \rangle $ and 
four particles in $|0,\pm 1 \rangle $. Thus, the configuration energy equals 
$2\hbar\omega + 4\cdot 2\hbar\omega $. This means that for the 
spectra with different cut-off energies displayed in 
Figure~\ref{f1} all excitations up to 
an energy $E_c - 10\hbar\omega $ are included.
The many-body spectra for $14\hbar \omega $ and 
$16\hbar \omega $ differ drastically
from the results obtained for $E_{c}\ge 20 \hbar\omega $.
Looking at the relative ordering of the levels and the spin sequence,
it becomes clear that 
convergence is only reached for $E_{c} > 20\hbar\omega $.
The ground-state energy for zero angular momentum is $E_0=3.049$~Ha$^*$
for $22\hbar \omega $ and $E_0=3.045$~Ha$^*$ for $24\hbar \omega $.
An extrapolation to infinite cut-off energy can be made by plotting 
the total energy as a function of 
$(E_{c}-10\hbar\omega )^{-3/2}$~\cite{koskinen2}.
This gives the estimate 3.043~Ha$^*$ 
for the fully converged results at $r_s=4~a_B^*$.
For $E_{c}=14\hbar \omega $ a too small number of Slater 
determinants was included to build up the required correlations, such that the 
polarized $S=3$ state appeared as the ground state.
(We identify a similar effect in unrestricted HF results mentioned 
above~\cite{landman}, where the single Slater determinant 
that is available incorrectly favors a spin-polarized ground 
state~\cite{pfann}.)
While for $E_{c}=22\hbar \omega $ the matrix dimension 
$44181$ with 21448811 non-zero matrix elements is reasonably small,
the value $E_{c}=24\hbar \omega $ already yields a matrix dimension 
108375 with 67521121 non-zero matrix elements.
As for larger densities the states are less correlated, a smaller number 
of Slater determinants is needed for an accurate description.
Density parameters larger than $r_s=4a_B^*$ or a higher number of 
particles than $N=6$ would go beyond the limits of numerical feasibility 
and accurate results could not be obtained.

\section{Many-body spectra of a six-electron quantum dot}

We now analyze the many-body spectra 
and the sequence of spins for the low-lying states as the 
two-dimensional density parameter $r_s$ is varied.
We choose the particle number $N=6$  as it corresponds to the 
smallest dot size for which classically two stable crystalline structures 
co-exist: A pentagonal ring with one electron at the center, and a slightly 
distorted six-fold ring~\cite{bolton1,bedanov}.  
We fix the angular momentum to $L=0$ and show in Figure~\ref{f2}
the 50 lowest states for a quantum dot confining $N=6$ particles 
at density parameters between $r_s=1~a_B^*$ and $r_s=4~a_B^*$
(in steps of $0.5~a_B^*$). 
To obtain a better resolution of the spectra 
the energies of the eigenstates $\epsilon _i$ are scaled such that 
the energy difference between the ground state and 50$^{th}$ excited 
state equals one, i.e. plotted are the dimensionless quantities
$\tilde \epsilon _i = (\epsilon _i -\epsilon _1 )/
(\epsilon _{50}-\epsilon _1 )$~.
(The total energies of the ground state with $S=0$ and 
the excited state with $S=3$ are given in Table~1 below.) 
At a very large density corresponding to $r_s=1~a_B^*$, 
the ground state has spin $S=0$ and is separated from the lowest 
excited state with spin $S=2$ by a gap of 
0.49~Ha$^*$.
This state is followed by a state with $S=1$ and again another (excited) 
spin singlet. The lowest fully polarized state is only found at a fairly 
high energy, the energy difference to the $S=0$ ground state being 
about 0.95~Ha$^*$.
(We note that for $N=6$ at $r_s=1.73~a_B^*$ we obtained excellent agreement 
of the $S=0$ ground state energy with the result of 
Pederiva~{\it et al.}~\cite{pederiva}.) 
As $r_s$ is increased, the fully polarized $S=3$ state moves down in energy.
At $r_s=2.5~a_B^*$ it has passed the excited singlet, but
is still far from competing with the nonpolarized $S=0$ ground state.
From the evolution of the energy difference of the $S=3$ state
to the ground state,
we do not expect any crossing of the polarized state and the ground state 
unless $r_s$ becomes {\it much} larger than $4~a_B^*$~.
Estimating the decrease in energy 
of the polarized state with respect to the ground state as 
$r_s$ is increased, our data seem to support the result of 
Egger~{\it et al.}~\cite{egger} that the ground state of the 6 electron dot 
is not polarized for $r_s$-values smaller than about 
$8~a_B^*$.
\PostScript{7}{0}{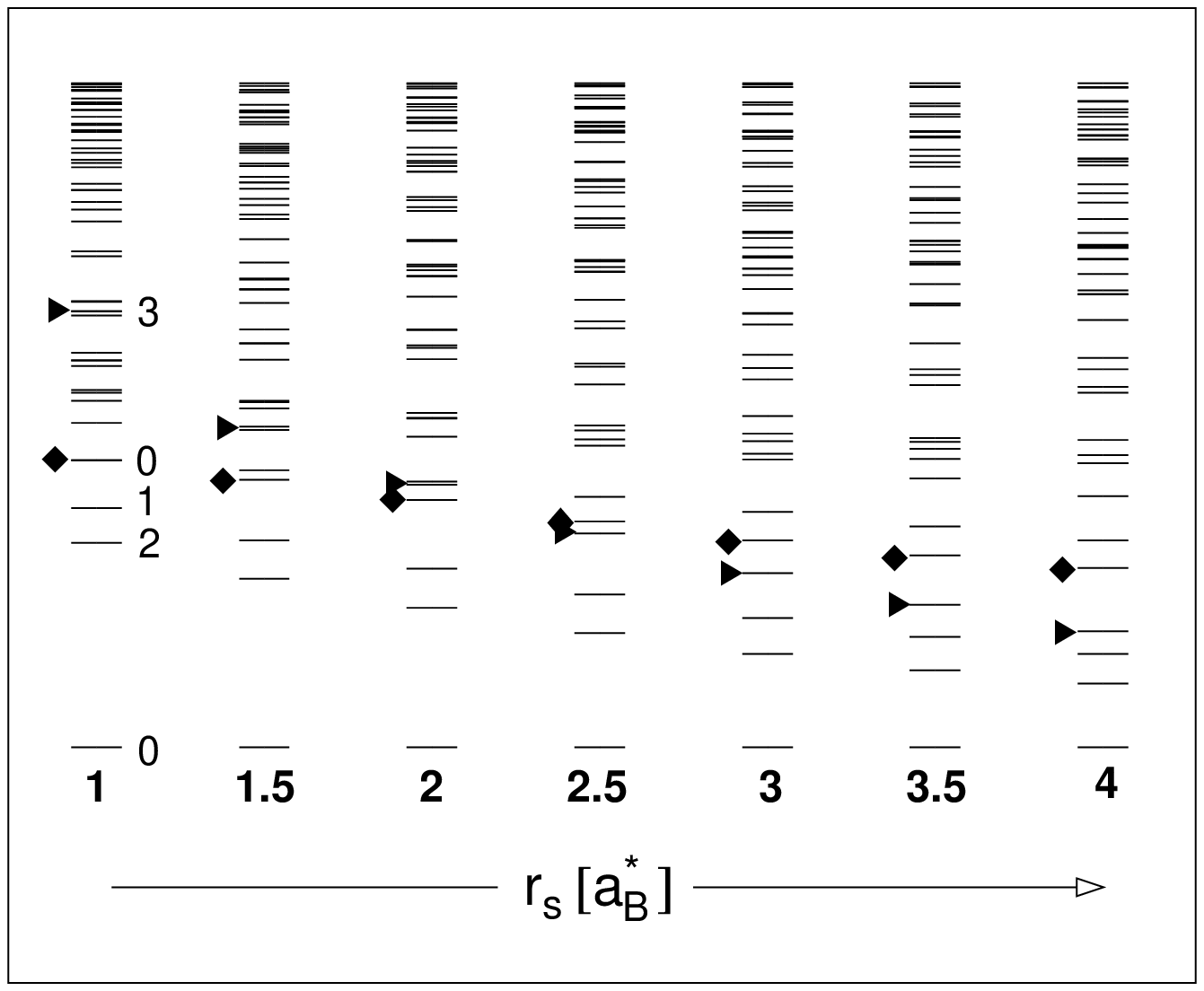}{0.5}{14}{\normalsize 
Spectra for angular momentum $L=0$ and six 
electrons in a harmonic trap at densities corresponding to 
Wigner-Seitz radii between $r_s=1~a_B^*$ and $r_s=4~a_B^*$ 
in steps of $0.5~a_B^*$. 
The square and the triangle show the first excited singlet state 
and the lowest fully polarized state, respectively.
The lowest state has always $S=0$.
The energy axis is scaled such that the energy difference between the 
ground state $\epsilon _1$ and 50$^{th}$ excited state $\epsilon _{50}$
equals one, i.e. plotted are the dimensionless quantities
$\tilde \epsilon _i = (\epsilon _i -\epsilon _1 )/
(\epsilon _{50}-\epsilon _1 )$~}{f2}
Table~1 compares the total energies of the ground state with spin zero and the 
lowest polarized state with spin $S=3$ with the corresponding 
result obtained from DFT,

\begin{center}
\bigskip
\begin{tabular}{lcccc}\hline\hline
 & \multicolumn{2}{l}{ ~~~~~paramagnetic~~~~~  } & 
\multicolumn{2}{l}{ ~~~~~ferromagnetic~~~~~ }\\ 
 $~r_s~[a_B^*]$~~~~ &~~exact~~ &~~LSDA~~ & ~~exact~~ &~~LSDA ~~\\ \hline
    &         &        &         &        \\
~1.0 & 14.27 & 14.30 & 15.22 & 15.30 \\
~1.5 & 8.983 & 8.988 & 9.363 & 9.409 \\ 
~2.0 & 6.508 & 6.503 & 6.695 & 6.724 \\
~2.5 & 5.084 & 5.073 & 5.188 & 5.204 \\
~3.0 & 4.162 & 4.148 & 4.225 & 4.233 \\
~3.5 & 3.519 & 3.502 & 3.559 & 3.560 \\
~4.0 & 3.045 & 3.027 & 3.071 & 3.068 \\ \hline\hline
\end{tabular} 
\end{center}
\medskip
{TABLE~1: Table of energies [Ha$^*$] for the 
paramagnetic $(S=0)$ and ferromagnetic
$(S=3)$ states in a 6-electron quantum dot for different densities. 
In the diagonalization we used 
$E_{c}=22\hbar \omega $ for $r_s\le 3.5~a_B^*$ and 
$E_{c}=24\hbar \omega $ for $r_s= 4~a_B^*$. For comparison the 
energies obtained with the local spin density approximation 
(LSDA) are also shown.}

\newpage
\noindent where the exchange-correlation part
of the electron-electron interactions is treated in LSDA.
For the DFT results we 
used an interpolation formula for the 
Tanatar-Ceperley~\cite{tanatar} exchange-correlation energy.
We refer to~\cite{koskinen} for the further details concerning the 
numerical method.
The CI energies of the $S=0$ ground states and $S=3$ isomer 
compare well with the LSDA results.
For the paramagnetic case, the LSDA gives lower energies than the exact 
results, when $r_s\ge~2~a_B^*$. This might be mainly due to the fact that
the Tanatar-Ceperley interpolation formula  slightly overestimates 
the correlation energy. Nevertheless, the LSDA gives 
surprisingly accurately the energy difference between the fully polarized ($S=3$)
and the paramagnetic ($S=0$) state, as seen in Fig.~\ref{f3}.
(For comparison, we also show the results for the infinite electron gas.)
\PostScript{9.5}{0}{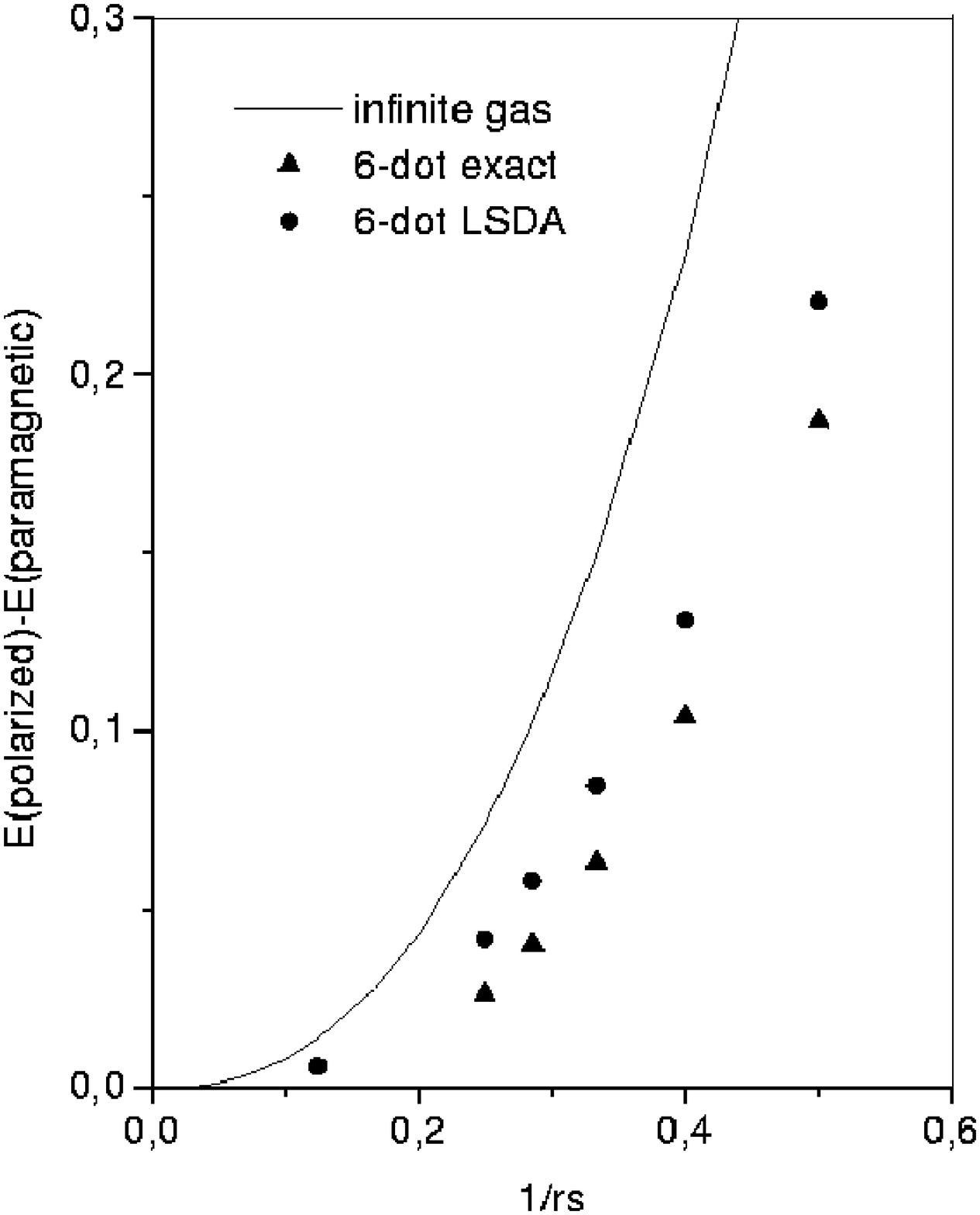}{0.5}{14}{\normalsize 
Energy between the fully polarized ($S=3$) and paramagnetic state ($S=0$)
for six electrons 
as a function of $1/r_s$. The result for the infinite 2D gas 
(per 6 electrons) is calculated 
using the interpolation formula of Tanatar and Ceperley~\cite{tanatar}.
}{f3}
Figure~\ref{f2} shows that for $r_s=~4~a_B^*$ there are only two
$L=0$ states between the ferromagnetic $(S=3)$ state and the 
paramagnetic $(S=0)$ ground state. However, if we consider 
all $L$-values, there are several states within this energy range.
This is shown in Table~2 where energies, spins and angular momenta
of all levels up to the first ferromagnetic state are given.
Indeed, the lowest excited state has $L=1$ and $S=1$. It is possible that 
at very large $r_s$ this partially polarized state might become the 
ground state instead of the fully polarized $S=3$ state~\cite{egger}.
\begin{center}
\bigskip
\begin{tabular}{llllll}\hline\hline
~~~~$r_s = 3a_B^*$ &~~ &~~ & $r_s = 4a_B^*$ &~~  &~~ \\
~~~~E [Ha$^*$]~~~~ & $S$~~ & $L$~~~~~~~~~ & E [Ha$^*$]~~~~ & $S$~~ & $L$ ~~~~\\ \hline
        &   &   &                 &     &     \\
~~~4.162   & 0 & 0 & 3.046 & 0 & 0 \\
~~~4.183   & 1 & 1 & 3.054 & 1 & 1 \\
~~~4.194   & 1 & 3 & 3.060 & 2 & 0 \\
~~~4.196   & 2 & 0 & 3.062 & 1 & 3 \\
~~~4.201   & 1 & 1 & 3.063 & 1 & 1 \\
~~~4.205   & 0 & 1 & 3.065 & 0 & 1 \\
~~~4.209   & 1 & 0 & 3.066 & 1 & 0 \\
~~~4.209   & 0 & 2 & 3.068 & 2 & 1 \\
~~~4.209   & 0 & 3 & 3.068 & 0 & 2 \\
~~~4.213   & 1 & 2 & 3.070 & 1 & 2 \\
~~~4.216   & 2 & 1 & 3.070 & 0 & 3 \\
~~~4.216   & 2 & 2 & 3.071 & 3 & 0 \\
~~~4.225   & 3 & 0 &       &   &   \\ \hline\hline
\end{tabular} 
\end{center}
\medskip
{TABLE~2: Table of energies [Ha$^*$], spins $S$ and angular momenta $L$
of all levels up to the lowest ferromagnetic state for $r_s=3a_B^*$
({\it left}) and $r_s=4a_B^*$ ({\it right}).
}
\section{Charge densities and pair correlation}

For $r_s=4~a_B^*$ the radial densities of the $S=0$ ground state and the 
3rd excited state of $L=0$, which is the lowest 
fully polarized state with spin $S=3$, 
are shown in Figure~\ref{f4}.
The azimuthally symmetric charge density for the polarized case 
shows a clear maximum at the center
surrounded by an outer ring of lower density. 
In the paramagnetic case, the density profile is more smooth,
the maximum density being at about $r\approx 6~a_B^*$.
The LSDA result shows a clear minimum at the origin, while
the exact result has a larger density at the center.
For comparison with the results of Egger~{\it et al.}~\cite{egger}
we also show the density $\rho (r)$ multiplied by a factor
$2\pi r$ (cf. lower panel of Fig.~\ref{f4}).
For the polarized state, 
the maximum at the center is now seen as a clear shoulder 
in the density profile. Note that this is missing in the 
paramagnetic ground state density. (This is in disagreement 
with the results of Egger~{\it et al.}~\cite{egger} who found for the 
ground state a density profile with a clear shoulder as in 
the polarized case.)
The azimuthal averages of the 
density profiles qualitatively have similarities with the 
broken symmetry solutions of the unrestricted HF~\cite{landman}
which for the paramagnetic case results in a ring of six electrons,
and for the ferromagnetic case (ground state in HF) a ring of five electrons  
with one electron in the center.
However, the localization of the electrons is largely exaggerated in HF. 
Opposite to HF,  the LSDA correctly gives the paramagnetic state 
as the ground state, and its
density profile resembles the exact result.
LSDA does not break the azimuthal symmetry until $r_s> 8a_B^*$ when 
spin- or charge density wave-like states can occur~\cite{koskinen}.
Purely classical~Monte Carlo~\cite{bedanov,bolton1} computations 
have shown that for $N<6$ the charges are distributed 
on the perimeter of the dot, and none of the particles occupies the 
dot center. 
This changes for $N=6$, where the 
charge distribution with lowest energy consists of 
five electrons sitting on a ring, with the remaining electron 
occupying the center of the dot. This configuration is labeled by $(5,1)$.
If all 6 particles are arranged on the dot perimeter (labeled by $(6,0)$), 
the classical state is stable but has a higher energy 
than the $(5,1)$-configuration.
\PostScript{6.5}{0}{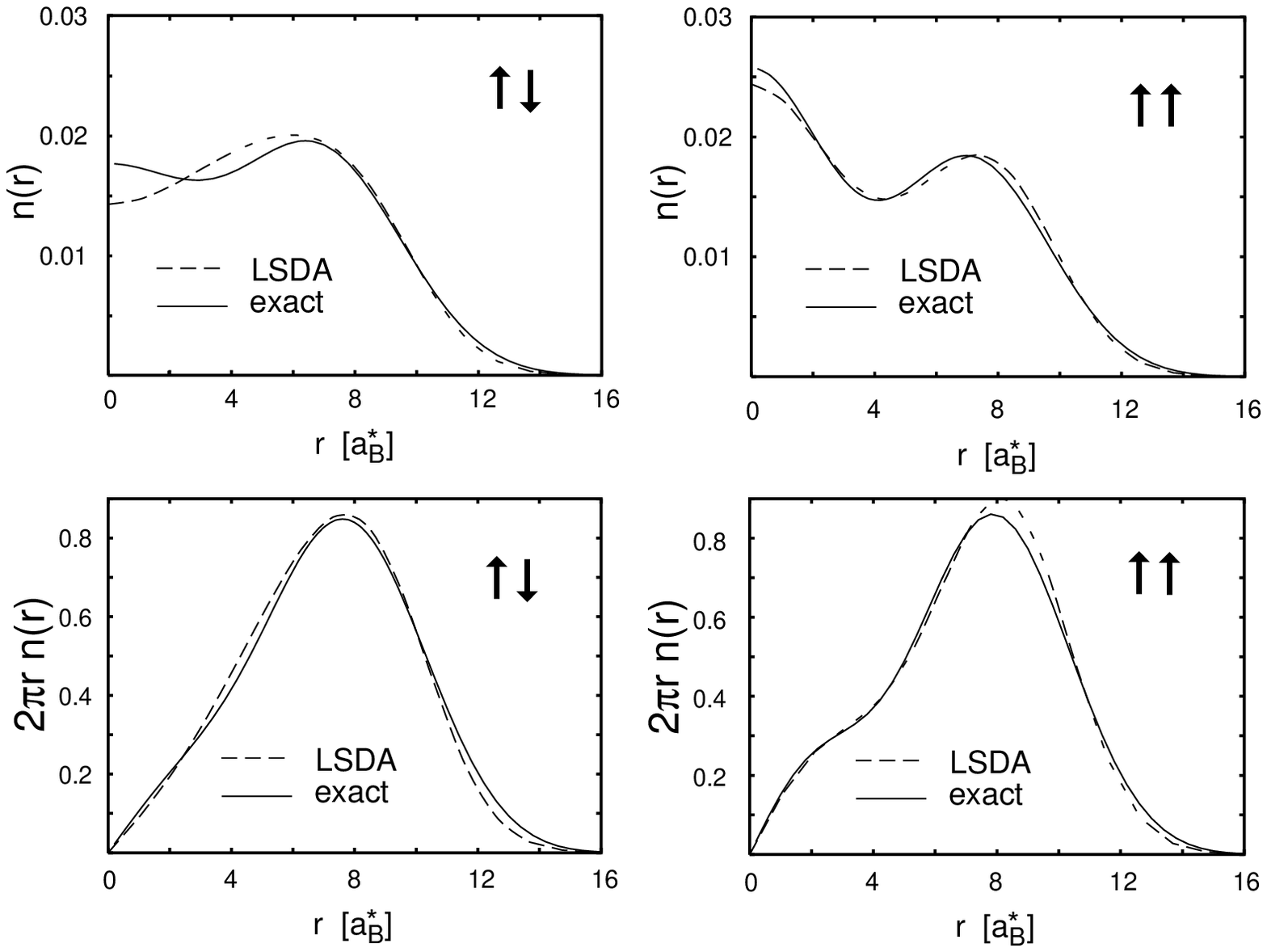}{0.5}{14}{\normalsize 
Charge density of a dot confining $N=6$ 
electrons in a harmonic trap at $r_s=4~a_B^*$; the exact result
({\it solid line}) is compared to the LSDA result ({\it dashed line}).
Shown is the density $n(r)$ {\it (upper panel)} and $n(r)$ multiplied by 
$2\pi r$ {\it (lower panel)} for the paramagnetic 
$S=0$ state {\it (left, $\uparrow\downarrow $)} and the ferromagnetic 
$S=3$ state {\it (right, $\uparrow\uparrow $)}.
The $S=3$ state is separated from the $S=0$ ground state by
0.026~Ha$^*$.}{f4}
The classical charge distribution can be arbitrarily oriented.
The density from the CI solution, however, must be circularly symmetric.
For an azimuthal average of the $(5,1)$-pentagon structure, one would expect
a pronounced maximum of the electron density in the center, and a 
less pronounced maximum at the dot radius.
Correspondingly, the $(6,0)$ configuration should correspond to 
a minimum of charge density in the center and a maximum at finite radius.
A first comparison of exact diagonalization calculations 
with the results of the mean-field approximation was given 
by Pfannkuche {\it et al.}~\cite{pfann}
for "quantum dot helium", i.e., quantum dots containing
only two electrons.
They found from a comparison of exact diagonalizations with
Hartree- and Hartree-Fock results that 
the exchange and correlation contributions are crucial. 
While the triplet state showed a reasonable agreement between 
the exact result and HF, the singlet could not be well
reproduced.
As mentioned above,
Yannouleas and Landman~\cite{landman} reported that in 
geometrically unrestricted HF at a density corresponding to 
$r_s\approx 3.5~a_B^*$ the $N=6$ ground state is polarized and 
shows enhanced localization in the charge density.
This $S=3$  state exhibits the 
same geometry than the classical distribution of 6 electrons in 
a harmonic well: 5 particles are equidistanly localized on the 
perimeter of the dot, and the 6th particle is trapped in the 
center of the harmonic well. 
The non-polarized $S=0$ state corresponding to the $(6,0)$-configuration
is about 0.034~Ha$^*$ higher in energy.
The exact diagonalization results described above do not support 
these HF results. Although being 
limited to small $r_s$ values due to the necessary restrictions 
of the basis set, the systematic evolution and energy sequence of 
CI energies and densities shown in Figs.~\ref{f2} and \ref{f4} 
seems to indicate that polarization as well as 
formation of Wigner molecules in 
circularly symmetric, parabolic wells would be impossible at 
densities as large as predicted by HF~\cite{landman}. 
The geometrically unrestricted solution of the Kohn-Sham equations 
of~\cite{koskinen} tend to overestimate the $r_s$-value at which 
spontaneously broken 
spin- or charge-symmetries can occur in the internal structure of the 
wave function~\cite{rings}. Although calculated in a geometrically 
unrestricted DFT scheme, the fully converged LSDA densities 
for $r_s \le 4 a_B^*$ shown in Fig.~\ref{f4} are azimuthally symmetric. 
Although LSDA suffers from the self-interaction problem, at the densities 
in question the results are in better agreement with CI studies than 
the unrestricted HF results.
\PostScript{8}{0}{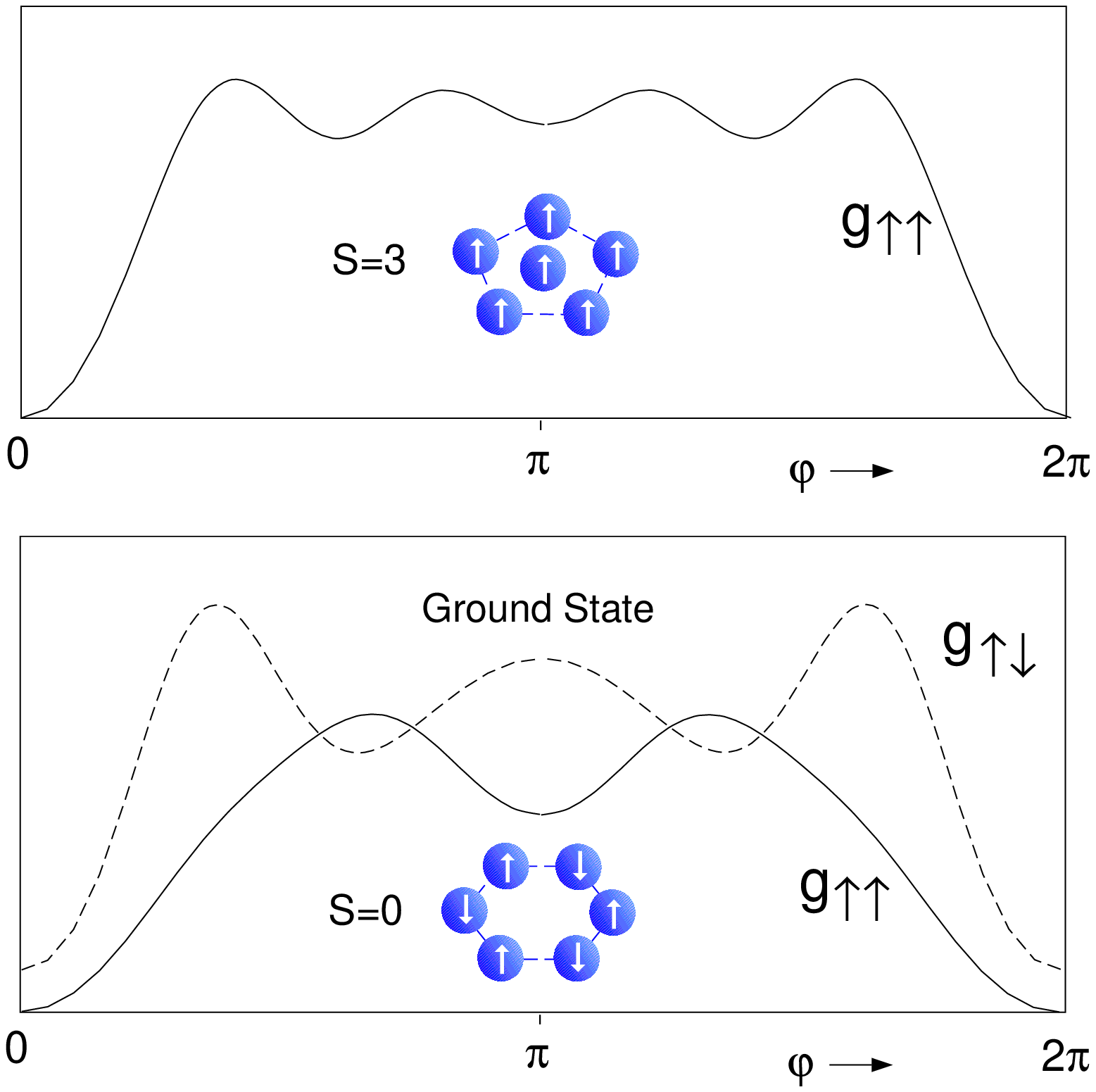}{0.5}{14}{\normalsize 
Pair correlation functions $g_{\uparrow\sigma}(\phi )$
calculated at the outer maxima of the density distribution (cf. 
upper panel in Fig.~\ref{f4}) as a function of the 
angle $\phi $ between the electrons for the ground state ({\it lower panel})  
and the excited polarized state with $S=3$ ({\it upper panel}).
The insets show schematic pictures of the electron configurations.
}{f5}
It is finally of interest to study how the spin and spatial symmetry in the 
internal structure of the wave function can be recognized in the pair 
correlation function
\begin{equation}
g_{\uparrow\sigma}(\varphi )=\langle \hat n_\uparrow(r,0)
\hat n_\sigma(r,\varphi ) \rangle~,
\end{equation}
which describes the probablity to find another (spin up or down) 
particle if a particle with spin up is placed at $(r,0)$. 
Here $r$ is the radius of maximum density and $\varphi $
the angle between the electrons. 
Figure~\ref{f5} shows  $g_{\uparrow\downarrow}(\varphi )$
and $g_{\uparrow\uparrow}(\varphi )$
for the ground state (lower panel)  
and $g_{\uparrow\uparrow}(\varphi )$ for 
the excited polarized state with $S=3$ (upper panel).
From the $\varphi $-values of the maxima in $g_{\uparrow\sigma}(\varphi )$
one clearly concludes that the $S=0$ state has 6-fold symmetry 
with antiferromagetic spin ordering, whereas the fully polarized 
case shows four maxima, corresponding to a five-fold symmetry.
These intrinsic symmetries are in qualitative agreement 
with the unrestricted HF results, although the crystallization predicted 
by HF does not yet occur.

\section{Conclusions}

We commented on the recent conjecture that Wigner molecules would form 
in quantum dots at rather large electron densities 
$r_s\stackrel{>}{\sim}~3.5~a_B^*$~\cite{landman,egger}.
Our results are essentially exact up to $r_s\le 4 a_B^*$
for six confined particles. 
The many-body spectra, densities and pair correlations 
obtained for $N=6$ clearly illustrate that 
the onset of formation of Wigner molecules 
and in particular, polarization of the ground state should be 
expected at much higher $r_s$-values than anticipated 
from unrestricted HF~\cite{landman}. The critical density 
at which such a transition occurs does, in fact, strongly 
depend also on geometry~\cite{creffield} and on the number of confined 
particles at fixed average electron density in the dot. 
For the six-electron dot at $r_s\approx 4a_B^*$ in question~\cite{landman},  
the ``exact'' ground-state clearly prefers $S=0$ and shows 
antiferromagnetic order in the pair correlation.
The polarized state with $C_{5v}$-symmetry is clearly higher in energy
even than the $S=1$ state, which would also be a candidate for 
the classical $(5,1)$ ground state configuration~\cite{bedanov,bolton1}.
In addition, for values below $r_s=4~a_B^*$ we did not find clear 
signals of rotational structure in the spectra for non-zero angular 
momenta that would indicate a crystallized ground state.
We note that the situation is different for $N<5$, where 
for densities as large as $r_s = 2 a_B^*$ the low-lying states 
could be well understood by assuming a square-shaped
$(4,0)$ Wigner molecule for the internal structure of the wave function
and analyzing its rotational structure. This also became clear when 
comparing the low-energy spectrum 
of a Heisenberg model with four electrons on a square~\cite{rings}.
For $N>5$, however,
this simple picture does seem to not hold, as our results 
for $N=6$ clearly point out.
The ground-state energies and densities 
obtained by density functional calculations  
in the local spin density approximation agree rather well with the results
of exact diagonalizations, even though this comparison was restricted to 
a small particle number where the accuracy of the local density 
approximation is questionable.
This gives some confidence that the method is well suited for 
describing the ground state electronic structures for larger sizes.

\bigskip

\bigskip

This work was supported by the Academy of Finland, 
the ``Bayerische Staatsministerium f\"ur 
Wissenschaft, Forschung und Kunst'' and 
the TMR programme of the European Community under contract 
ERBFMBICT972405.

\end{multicols}
\end{document}